\documentclass[aps,prapplied,preprint,superscriptaddress,amsmath,amssymb,floatfix]{revtex4-2}

\usepackage{graphicx}
\usepackage{xcolor}
\usepackage{soul}
\usepackage{siunitx}
\sethlcolor{yellow}
\soulregister{\cite}{1}
\soulregister{\ref}{1}
\soulregister{\eqref}{1}
\DeclareRobustCommand{\rev}[1]{#1}
\newcommand{\revm}[1]{#1}

\begin{document}

\title{Parameterized Physics-Informed Neural Networks for Efficient Simulation of Embedded Microfluidic Chip Cooling}

\author{Weiheng Li}
\affiliation{School of Mechanics and Engineering Science, Peking University, Beijing 100871, China}

\author{Zhiyao Jiang}
\affiliation{School of Mechanics and Engineering Science, Peking University, Beijing 100871, China}

\author{Bai Song}
\email{songbai@pku.edu.cn}
\affiliation{School of Mechanics and Engineering Science, Peking University, Beijing 100871, China}
\affiliation{National Key Laboratory of Advanced Micro and Nano Manufacture Technology, Beijing 100871, China}

\date{\today}

\begin{abstract}
Embedded microfluidic cooling is a promising technology for dissipating the intense heat generated by high-performance electronic chips. However, the design and optimization of such coolers impose prohibitive computational costs when using conventional numerical solvers. Here, we present a parameterized physics-informed neural network (PINN) as a deep learning framework for rapidly and accurately predicting the thermohydraulic performance of embedded manifold microchannels. By incorporating four geometric dimensions along with three spatial coordinates as the network inputs, the proposed model enables multiphysics prediction across a continuous design space. Remarkably, our predicted velocity and temperature fields differ from finite-element simulations by \rev{relative errors of only 8.1\% and 5.6\%}, respectively, while reducing the computational time and cost by over four orders of magnitude. Key to this framework is a composite strategy combining geometric ansatz embedding, the SOAP optimizer, and a thermal-conductivity-based curriculum learning scheme, which helps overcome the convergence challenges inherent in training PINNs for complex three-dimensional flows. Notably, by leveraging a highly optimized JAX backend, the network achieves convergence on a single consumer-grade GPU in less than four hours. Our approach offers a physics-consistent, real-time surrogate model that can substantially accelerate innovations in advanced microfluidic cooling of future electronics.
\end{abstract}


\maketitle

\section{Introduction}
\label{sec1}

The continuous escalation of power density in integrated circuits necessitates more advanced and efficient thermal dissipation strategies \cite{fabbri_state_2007, tuckerman_high-performance_1981}. For instance, in high-performance computing, microprocessors routinely generate heat fluxes that exceed the capacity of conventional air cooling \cite{zhang_survey_2021}. Moreover, radio-frequency and power electronics based on wide-bandgap semiconductors such as silicon carbide and gallium nitride operate at much higher power densities \cite{bar-cohen_embedded_2019}. Without adequate cooling, elevated junction temperatures degrade device performance, reduce reliability, and shorten operational lifespan \cite{bar-cohen_icecool_2021}. Consequently, there has been a marked shift toward advanced liquid cooling research to address these escalating thermal demands \cite{fabbri_state_2007, van_erp_co-designing_2020}.

Among diverse approaches, embedded microfluidic cooling has emerged as a promising solution for dissipating ultrahigh heat fluxes \cite{escher_experimental_2010, van_erp_co-designing_2020, wu_jet-enhanced_2025}. Unlike conventional cold plates which are mounted on the chip surface via a layer of thermal interfacial material, embedded cooling leverages microfluidic structures that are etched directly into the backside of a chip. This approach brings the coolant as close as possible to the hotspots, which fundamentally reduces the thermal resistance and therefore enables unprecedented cooling performance. In particular, manifold microchannel configurations offer unique advantages, which help maintain a low pressure drop while substantially enhancing heat transfer \cite{lin_single-phase_2021}. Remarkably, recent studies have numerically predicted and experimentally demonstrated that embedded manifold microchannels (EMMC) with single-phase water can dissipate heat fluxes exceeding \qty{3000}{\watt\per\centi\meter\squared} over a heated area of \qty{9}{\milli\meter\squared} \cite{wu_jet_2023, wu_efficient_2024, wu_jet-enhanced_2025}. \rev{In addition, embedded cooling with a dielectric fluid has also achieved record-high heat fluxes up to \mbox{\qty{1570}{\watt\per\centi\meter\squared}} \mbox{\cite{xiao_horizontal-up_2026}}}. Of key importance is the geometric parameters of these structures which strongly impact the thermal and hydraulic performance \rev{\mbox{\cite{jiang_synergistic_2026, jiang_embedded_2026, li_biomimetic_2026}}}. Therefore, tailored design and optimization for a specific chip becomes necessary when moving towards real-world applications. However, parametric optimization of EMMC requires a great number of high-fidelity simulations of three-dimensional (3D) conjugate heat transfer, which is expensive especially when using conventional approaches such as the finite-element method (FEM). When exploring a design space with multiple geometric parameters, the total computational cost becomes prohibitive \cite{jiang_synergistic_2026, yang_parameterized_2025}.

\rev{To accelerate thermohydraulic simulation and reduce the cost of optimization, machine learning (ML) has emerged as a promising framework. Recently, data-driven ML approaches have been widely employed in thermal science, particularly for the prediction of thermophysical properties and the design of thermal functional materials and devices \mbox{\cite{yu_general_2023, hu_machine-learning-optimized_2020, hu_machine_2020, yu_ultrahigh_2024, chen_strong_2025, zhou_insight_2025, wu_correcting_2024}}. Furthermore, automated data analysis for laser pump-probe thermal measurement has been demonstrated \mbox{\cite{yang_vibe-fdtr_2026}}. Complementary to various data-driven methods, physics-informed neural networks (PINNs) \mbox{\cite{raissi_physics-informed_2019}}stand as an appealing alternative for fluid flow and heat transfer modeling}. PINNs incorporate the governing partial differential equations (PDEs) directly into the loss function, enabling the network to maintain physical consistency and learn solutions with little or even no labeled training data \cite{raissi_physics-informed_2019, cuomo_scientific_2022}. Given their diverse advantages, PINNs have been applied to various problems in fluid mechanics and heat transfer \cite{raissi_hidden_2020, hennigh_nvidia_2021, cuomo_scientific_2022, zhang_physics-informed_2024, yang_parameterized_2025, wang_simulating_2025}. Notably, Hennigh et al. \cite{hennigh_nvidia_2021} pioneered parameterized PINN simulations for heat sinks using the NVIDIA SimNet™ framework (currently known as PhysicsNeMo Sym \cite{noauthor_physicsnemo_nodate}). Zhang et al. utilized PINNs to simulate conjugate heat transfer in manifold microchannel heat sinks, focusing on modeling for a fixed geometry \cite{zhang_physics-informed_2024}. Yang et al. proposed a parameterized PINN for the thermal design of a power module based on a layer of millimeter-scale pin-fins \cite{yang_parameterized_2025}. Despite such progress, it remains a great challenge to apply PINNs to complex 3D conjugate heat transfer with parameterized geometries. First, the computational cost of training 3D PINNs remains high \cite{zhang_physics-informed_2024, hennigh_nvidia_2021}. Further, the training of PINNs for convection-dominated heat transfer faces a variety of optimization difficulties \cite{wang_experts_2023, krishnapriyan_characterizing_2021}. Finally, parameterized geometries introduce additional complexity that as design parameters change, the spatial relationship between sampling points and domain boundaries varies, making it difficult for networks to distinguish boundaries from interior regions.

Here, we present a parameterized PINN for the 3D simulation of fluid flow and heat transfer in embedded jet-enhanced manifold microchannels, overcoming the aforementioned challenges through several methodological contributions. First, a low cost geometric ansatz embedding strategy encodes the relative spatial position of each point within the parameterized geometry, providing explicit geometric information essential for convergence. Second, a curriculum learning scheme \cite{bengio_curriculum_2009} transitions training from diffusion-dominated to convection-dominated regimes by gradually adjusting the effective Prandtl number and the fluid--solid conductivity ratio. This mitigates instabilities common in PINN training for high-Prandtl-number flows. Third, the SOAP optimizer \cite{vyas_soap_2025} combined with gradient normalization \cite{wang_experts_2023} enables stable convergence in this multiphysics problem. Fourth, the implementation uses a framework based on JAX \cite{bradbury_jax_2018, deepmind_deepmind_2020} with forward-mode automatic differentiation \cite{baydin_automatic_2018, li_computational_2024}, achieving efficient training on a single consumer-grade GPU in only four hours. Our network accepts the three spatial coordinates along with four geometric parameters as inputs, including microchannel height, width, and jet inlet/outlet widths. Once trained, the model predicts the coupled flow and temperature fields for any parameter combination in a continuous design space within milliseconds, enabling over four orders of magnitude reduction in per-design computational time and cost relative to conventional finite-element simulations on a 56-core workstation. Validation against commercial FEM software shows \rev{relative errors of 8.1\% and 5.6\%} for the velocity and temperature fields, respectively. The one-time training cost is comparable in wall-clock time to only 108 single-design FEM simulations, a level that is readily offset in parameter sweeps and optimization studies involving larger numbers of design evaluations. This physics-consistent surrogate model can therefore substantially accelerate the design and optimization of embedded microfluidic coolers.

\section{Methods}
\label{sec2}

\subsection{Physical model and geometric parameterization}

\begin{figure}
    \centering
    \includegraphics[width=\linewidth]{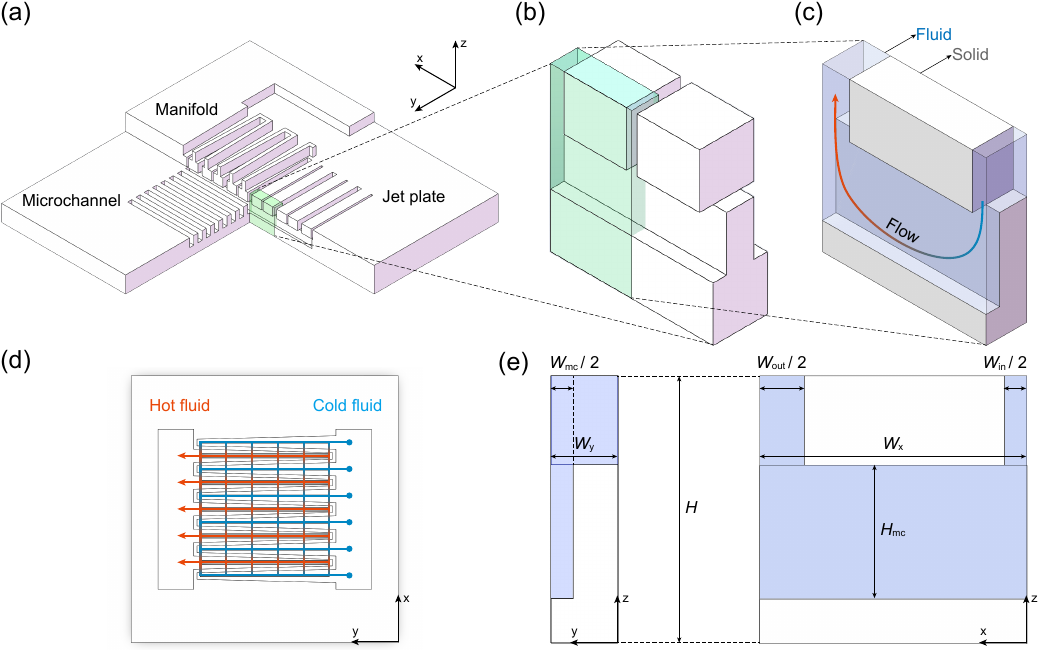}
    \caption{\textbf{Geometric configuration and computational domain of the jet-enhanced manifold microchannel cooler.} \textbf{(a)} 3D schematic of the three-tier cooling structure, comprising the manifold, jet plate, and microfluidic layers. \textbf{(b)} A representative unit extracted from the full device as indicated by the green shade in (a). \textbf{(c)} The modeled computational domain which is symmetry-reduced and corresponds to the green shade in (b). The blue and white regions represent the fluid (water) and solid (silicon) domains, respectively, with the colored arrow showing the fluid flow from the jet inlet to the outlet. \textbf{(d)} Top-view schematic illustrating the alternating flow layout of cold fluid (inlets) and hot fluid (outlets) within the manifold layer. \textbf{(e)}~Cross-sectional projections defining the key geometric dimensions. $H$, $W_x$, and $W_y$ denote the overall outer dimensions of the domain. Due to the applied symmetric conditions, the physical widths of the microchannel, inlet, and outlet are depicted as half-widths. Notably, $W_\text{in}$, $W_\text{out}$, $W_\text{mc}$, and $H_\text{mc}$ constitute the four geometric parameters incorporated into the PINN to form a continuous design space.}
    \label{fig:geometry}
\end{figure}

In this study, we employ an embedded jet-enhanced manifold microchannel cooler as a representative complex three-dimensional conjugate heat-transfer system consisting of solid and fluid regions. As shown in Fig.~\ref{fig:geometry}(a)-(c), the computational domain is symmetry-reduced from the entire cooler to a single microchannel unit, with symmetry boundary conditions imposed on the corresponding planes to minimize cost. The solid domain is the substrate that conducts heat from the active device layer to the coolant, while the fluid domain represents the coolant passages. Without loss of generality, we consider here a silicon substrate in conjunction with water as the coolant.

To define the parameterized design space, four key geometric dimensions are incorporated including the jet inlet width $(W_{\mathrm{in}} \in [40,50]~\unit{\micro\meter})$, the jet outlet width $(W_{\mathrm{out}} \in [90,110]~\unit{\micro\meter})$, the microchannel width $(W_{\mathrm{mc}} \in [45,55]~\unit{\micro\meter})$, and the microchannel height $(H_{\mathrm{mc}} \in [140,160]~\unit{\micro\meter})$. These parameters are selected based on prior design studies in which they were found to have a strong influence on the thermohydraulic performance of the device \cite{jiang_synergistic_2026}. Their ranges are chosen to remain compatible with practical manufacturing constraints while covering a sufficiently large design space.

Steady-state heat transfer in the EMMC structure is governed by the incompressible Navier--Stokes and energy equations in the fluid domain, and by the heat conduction equation in the solid domain. The 3D space is captured by the Cartesian coordinate $\mathbf{r}=[x,y,z]$. In the fluid domain, conservation of mass requires
\begin{equation}
\nabla \cdot \mathbf{u} = 0,
\end{equation}
where $\mathbf{u}=(u,v,w)$ denotes the velocity vector. Conservation of momentum is expressed as
\begin{equation}
(\mathbf{u}\cdot\nabla)\mathbf{u} = -\frac{1}{\rho}\nabla p + \nu \nabla^2 \mathbf{u},
\end{equation}
where $p$ is the pressure, $\rho$ is the fluid density, and $\nu$ is the kinematic viscosity. The fluid energy equation is
\begin{equation}
(\mathbf{u}\cdot\nabla)T_\mathrm{f} = \alpha_\mathrm{f} \nabla^2 T_\mathrm{f},
\end{equation}
where $T_\mathrm{f}$ is the fluid temperature and $\alpha_\mathrm{f}$ is the thermal diffusivity of the fluid. In the solid domain, heat conduction is governed by
\begin{equation}
\nabla^2 T_\mathrm{s} = 0,
\end{equation}
where $T_\mathrm{s}$ is the solid temperature.

At the fluid-solid interface, temperature heat-flux continuity are imposed. Other conditions include the inlet and outlet boundaries in the fluid domain, the isothermal base and adiabatic top boundary in the solid domain, and also the symmetry planes. All thermophysical properties are treated as constants for simplicity, therefore the momentum equations are independent of temperature, while the energy equation depends on the velocity field.

The initial parameterized input vector is defined as
\begin{equation}
\mathbf{x}_{\mathrm{input}} = [x, y, z, W_{\mathrm{in}}, W_{\mathrm{out}}, W_{\mathrm{mc}}, H_{\mathrm{mc}}].
\end{equation}
This formulation shifts the learning target from a single fixed geometry to a family of solutions parameterized by geometry.

\subsection{Parameterized PINN framework}

A parameterized physics-informed neural network (PINN) is used to learn the mapping from spatial coordinates and geometric parameters to the corresponding physical fields. Specifically, a neural network $NN_{\theta}$ parameterized by weights and biases $\theta$ is constructed to approximate the state vector as
\begin{equation}
[u,v,w,p,T_\mathrm{f},T_\mathrm{s}] = NN_{\theta}(\mathbf{x}_{\mathrm{aug}}),
\end{equation}
where $\mathbf{x}_{\mathrm{aug}}$ denotes the augmented network input. The network is trained by minimizing a composite loss that includes the governing-equation residuals and the boundary-condition terms.

The network backbone is based on PirateNet \cite{wang_piratenets_2024}. To improve the representation of multiscale features, the input is first mapped into a higher-dimensional space using Fourier feature embeddings \cite{rahaman_spectral_2019},
\begin{equation}
\gamma(\mathbf{x}_\mathrm{aug}) =
\left[
\cos\left(2\pi \mathbf{W}_{\mathrm{emb}}\mathbf{x}_\mathrm{aug}\right),
\sin\left(2\pi \mathbf{W}_{\mathrm{emb}}\mathbf{x}_\mathrm{aug}\right)
\right],
\end{equation}
where $\mathbf{W}_{\mathrm{emb}} \in \mathbb{R}^{d\times m}$ is a learnable weight matrix and $m$ is the embedding dimension, set to 256 in this study. The final network consists of four PirateNet residual blocks, each with 256 hidden units. Random weight factorization is additionally used to facilitate optimization \cite{wang_random_2022}.

\begin{figure}
    \centering
    \includegraphics[width=\linewidth]{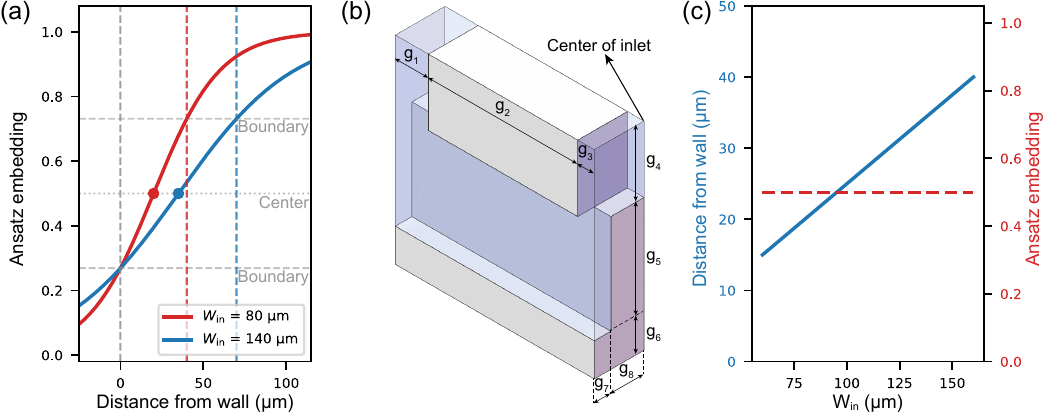}
    \caption{\textbf{Spatial layout and relative position invariance of the geometric ansatz embeddings.} \textbf{(a)} Mapping profiles from physical distances to the unified ansatz embedding space \rev{under two illustrative parametric conditions (chosen for visual contrast, whereas the training range is defined in Section~\mbox{\ref{sec2}}).} Despite variations in absolute physical dimensions, this mapping ensures that relative geometric features, such as boundaries or centers, consistently correspond to identical embedding values. For instance, the geometric center always aligns exactly with an embedding value of 0.5. \textbf{(b)} 3D spatial layout illustrating the placement of the eight designated geometric ansatz embeddings, denoted as $g_1$ through $g_8$, within the computational domain. \textbf{(c)} Validation of the invariance property at the center of the inlet. As the geometric parameter $W_\text{in}$ varies, the actual physical distance from the wall increases linearly as indicated by the blue solid line. Meanwhile, the corresponding ansatz embedding value remains constant as indicated by the red dashed line.}
    \label{fig:ansatz}
\end{figure}

To better represent the parameterized geometry, the initial input $\mathbf{x}_{\mathrm{input}}$ is further augmented with geometric ansatz embeddings. Variations in the geometric parameters can change both the domain to which a point belongs and its relative position with respect to nearby boundaries. To encode this information explicitly, normalized local coordinates are defined for each coordinate direction within the distinct domains:
\begin{equation}
\mathbf{g}_i = \frac{\mathbf{r}_i - \mathbf{r}_{i,\min}(\mathbf{p})}
{\mathbf{r}_{i,\max}(\mathbf{p}) - \mathbf{r}_{i,\min}(\mathbf{p})},
\end{equation}
where $\mathbf{p}=[W_{\mathrm{in}}, W_{\mathrm{out}}, W_{\mathrm{mc}}, H_{\mathrm{mc}}]$ is the current vector of geometric parameters and $\mathbf{r}=[x,y,z]$ is the spatial coordinate. These local coordinates are then transformed by a smooth embedding function. In this work, a sigmoid function $\sigma_\mathrm{s}$ is adopted,
\begin{equation}
\sigma_s(x) = \frac{1}{1+e^{-sx}},
\end{equation}
and the mapped ansatz feature is defined as
\begin{equation}
\tilde{\mathbf{g}}_i = \sigma_s\left(\frac{\mathbf{g}_i - 0.5}{\delta}\right),
\end{equation}
where $\delta$ is a scaling parameter controlling the sharpness of the transition. The augmented network input is then written as
\begin{equation}
\mathbf{x}_{\mathrm{aug}} = [\mathbf{x}_{\mathrm{norm}}, \tilde{\mathbf{g}}],
\end{equation}
where \(\mathbf{x}_{\mathrm{norm}}\) comprises seven normalized spatial coordinates and geometric parameters, and \(\tilde{\mathbf{g}}\) denotes eight mapped local geometric features, which together constitute a fifteen-dimensional network input in the present implementation, as illustrated in Figure~\ref{fig:ansatz}(b).

Training points are generated on the fly in JAX \cite{bradbury_jax_2018} rather than pre-generated externally. The three-dimensional fluid and solid domains are algebraically decomposed into parameterized bounding boxes for interior regions and boundary planes for surface constraints. Interior points are sampled uniformly over the composite multi-block regions corresponding to the jet inlet, microchannels, and jet outlet, whereas boundary points are sampled directly on the corresponding parameterized surfaces. A region-aware importance sampling strategy is further used so that more points are placed near the fluid-solid walls, narrow microchannel gaps, and inlet/outlet boundaries, where velocity and temperature gradients are strongest. Because the sampling is performed directly on a GPU, repeated CPU-to-GPU data transfer is avoided during training, thereby substantially improving performance.

\subsection{Loss formulation}

To improve numerical conditioning, all governing equations and boundary conditions are cast into a nondimensional form. The characteristic scales used in this work are $L_{\mathrm{ref}} = \qty{100}{\micro\meter}$ for the length,  $U_{\mathrm{ref}} = \qty{1}{\meter\per\second}$ for the velocity, $P_{\mathrm{ref}} = \rho U_{\mathrm{ref}}^2$ for the pressure, and $T_{\mathrm{ref}} = \qty{50}{\kelvin}$ for the temperature. The corresponding nondimensional variables are defined as
\begin{equation}
\mathbf{x}^* = \frac{\mathbf{x}}{L_{\mathrm{ref}}}, \qquad
\mathbf{u}^* = \frac{\mathbf{u}}{U_{\mathrm{ref}}}, \qquad
p^* = \frac{p}{\rho U_{\mathrm{ref}}^2}, \qquad
T^* = \frac{\Delta T}{T_{\mathrm{ref}}}.
\end{equation}

The fluid continuity and momentum equations become
\begin{equation}
\nabla^* \cdot \mathbf{u}^* = 0,
\end{equation}
\begin{equation}
(\mathbf{u}^* \cdot \nabla^*)\mathbf{u}^*
=
-\nabla^* p^* + \frac{1}{Re}\nabla^{*2}\mathbf{u}^*,
\end{equation}
while the energy equations in the fluid and solid domains are written respectively as
\begin{equation}
(\mathbf{u}^* \cdot \nabla^*)T_f^*
=
\frac{1}{Re\cdot Pr}\nabla^{*2}T_f^*,
\label{eq:energy_f}
\end{equation}
\begin{equation}
\nabla^{*2}T_s^* = 0.
\end{equation}
Here,
\begin{equation}
Re = \frac{U_{\mathrm{ref}}L_{\mathrm{ref}}}{\nu}, \qquad
Pr = \frac{\nu}{\alpha_f},
\end{equation}
denote the Reynolds number and Prandtl number, respectively. At the fluid--solid interface, continuity of heat flux gives
\begin{equation}
\frac{\partial T_f^*}{\partial n^*}
=
k^* \frac{\partial T_s^*}{\partial n^*},
\end{equation}
where
\begin{equation}
k^* = \frac{k_s}{k_f}
\end{equation}
denotes the thermal conductivity ratio.

To keep the network outputs within prescribed ranges, a linear output scaling is appended to the final layer:
\begin{equation}
[u^*,v^*,w^*,p^*,T_f^*,T_s^*]
=
\frac{NN_{\theta}(\mathbf{x}_{\mathrm{aug}})+1}{2}
\cdot
(\mathbf{y}_{\max}-\mathbf{y}_{\min})
+
\mathbf{y}_{\min},
\end{equation}
where $\mathbf{y}_{\min}$ and $\mathbf{y}_{\max}$ are empirical lower and upper bounds for the output variables.

The interior PDE losses are defined as the mean-squared residuals of the governing equations evaluated at dynamically sampled points. For the fluid domain, the continuity and momentum residuals are
\begin{equation}
\mathcal{L}_{\mathrm{cont}}
=
\frac{1}{N_f}
\sum_{i=1}^{N_f}
\left(\nabla^*\cdot\mathbf{u}^*\right)^2,
\end{equation}
\begin{equation}
\mathcal{L}_{\mathrm{mom}}
=
\frac{1}{N_f}
\sum_{i=1}^{N_f}
\left\|
(\mathbf{u}^* \cdot \nabla^*)\mathbf{u}^*
+
\nabla^* p^*
-
\frac{1}{Re}\nabla^{*2}\mathbf{u}^*
\right\|^2.
\end{equation}

The energy residuals in the fluid and solid domains are
\begin{equation}
\mathcal{L}_{\mathrm{en},f}
=
\frac{1}{N_f}
\sum_{i=1}^{N_f}
\left[
(\mathbf{u}^* \cdot \nabla^*)T_f^*
-
\frac{1}{Re\cdot Pr}\nabla^{*2}T_f^*
\right]^2,
\end{equation}
\begin{equation}
\mathcal{L}_{\mathrm{en},s}
=
\frac{1}{N_s}
\sum_{i=1}^{N_s}
\left(\nabla^{*2}T_s^*\right)^2.
\end{equation}

Boundary and interface conditions are enforced through additional loss terms. For the inlet boundary, the hydraulic and thermal constraints are defined separately because the thermal part is introduced progressively during training:
\begin{equation}
\mathcal{L}_{\mathrm{inlet,scalar}}
=
\frac{1}{N_{\mathrm{in},f}}
\sum_{i=1}^{N_{\mathrm{in},f}}
\left[
(p^* - p_{\mathrm{in}}^*)^2
+
(u^*)^2
+
(v^*)^2
\right],
\end{equation}
\begin{equation}
\mathcal{L}_{\mathrm{inlet,flux}}
=
\frac{1}{N_{\mathrm{in},f}}
\sum_{i=1}^{N_{\mathrm{in},f}}
\left(
\frac{1}{\revm{Re\cdot Pr}}\frac{\partial T_f^*}{\partial n^*}
-
w^* T_f^*
\right)^2,
\end{equation}
\begin{equation}
\mathcal{L}_{\mathrm{inlet,temp}}
=
\frac{1}{N_{\mathrm{in},f}}
\sum_{i=1}^{N_{\mathrm{in},f}}
\left(T_f^*\right)^2
+
\frac{1}{N_{\mathrm{in},s}}
\sum_{i=1}^{N_{\mathrm{in},s}}
\left(
T_s^* - T_{\mathrm{aux}}^*
\right)^2,
\end{equation}
where $T_{\mathrm{aux}}^*=0.8$ is an auxiliary non-dimensional temperature target used only during the early stage of training. The value is chosen empirically as a numerically convenient thermal anchor rather than a physically exact constraint. Since the weight of this auxiliary term decays to zero during training, it does not affect the final boundary condition enforced by the model.

At the outlet boundary,
\begin{equation}
\mathcal{L}_{\mathrm{outlet}}
=
\frac{1}{N_{\mathrm{out}}}
\sum_{i=1}^{N_{\mathrm{out}}}
\left[
(p^*)^2
+
(u^*)^2
+
(v^*)^2
+
\left(
\frac{\partial T_f^*}{\partial n^*}
\right)^2
\right].
\end{equation}

For the heated base of the solid domain,
\begin{equation}
\mathcal{L}_{\mathrm{base},s}
=
\frac{1}{N_{\mathrm{base},s}}
\sum_{i=1}^{N_{\mathrm{base},s}}
\left(T_s^* - T_{\mathrm{base}}^*\right)^2.
\end{equation}

For the solid top boundary,
\begin{equation}
\mathcal{L}_{\mathrm{top}}
=
\frac{1}{N_{\mathrm{top}}}
\sum_{i=1}^{N_{\mathrm{top}}}
\left(
\frac{\partial T_s^*}{\partial n^*}
\right)^2.
\label{eq:top_loss}
\end{equation}

At the fluid-solid interface, no-slip, temperature continuity, and heat-flux continuity are imposed through
\begin{equation}
\mathcal{L}_{\mathrm{noslip}}
=
\frac{1}{N_w}
\sum_{i=1}^{N_w}
\|\mathbf{u}^*\|^2,
\end{equation}
\begin{equation}
\mathcal{L}_{\mathrm{temp,cont}}
=
\frac{1}{N_w}
\sum_{i=1}^{N_w}
\left(T_f^* - T_s^*\right)^2,
\end{equation}
\begin{equation}
\mathcal{L}_{\mathrm{flux,cont}}
=
\frac{1}{N_w}
\sum_{i=1}^{N_w}
\left(
\frac{\partial T_f^*}{\partial n^*}
-
k^*\frac{\partial T_s^*}{\partial n^*}
\right)^2.
\end{equation}

For the symmetry planes, the velocity and scalar losses are defined as
\begin{equation}
\mathcal{L}_{\mathrm{sym},u}
=
\frac{1}{N_{\mathrm{sym},f}}
\sum_{i=1}^{N_{\mathrm{sym},f}}
\left[
(\mathbf{u}^*\cdot\mathbf{n})^2
+
\revm{\left\|(\mathbf{I}-\mathbf{n}\mathbf{n}^{T})\frac{\partial \mathbf{u}^*}{\partial n^*}\right\|^2}
\right],
\end{equation}
\begin{equation}
\mathcal{L}_{\mathrm{sym},\phi}
=
\frac{1}{N_{\mathrm{sym}}}
\sum_{i=1}^{N_{\mathrm{sym}}}
\left(
\frac{\partial \phi}{\partial n^*}
\right)^2,
\qquad
\phi \in \{p^*,T_f^*,T_s^*\}.
\end{equation}

\subsection{Training strategy and implementation}
\label{sec2_4}
Curriculum learning is employed to enhance the robustness of PINN training for the present conjugate heat-transfer problem. Such a strategy is necessary because the target regime is convection dominated and involves a strong thermal-conductivity contrast across the fluid-solid interface, which significantly increases the training difficulty. To this end, a unified continuation factor is introduced:
\begin{equation}
\alpha(t)=\min\left(\frac{t}{t_{\mathrm{warmup}}},1\right).
\end{equation}
\rev{Here, \mbox{$t_{\mathrm{warmup}}=25,000$}, which is one eighth of the total 200,000 training steps. This factor is used consistently to control the thermal curriculum and activation of the heated-base thermal constraint}.

The effective Prandtl number used during training is scheduled as
\begin{equation}
Pr_{\mathrm{eff}}(t)
=
Pr_{\min}
+
\left(Pr_{\mathrm{target}} - Pr_{\min}\right)\alpha(t),
\end{equation}
where $Pr_{\min}=10^{-6}$ and $Pr_{\mathrm{target}}=6.977$. At the early stage of training, the small effective Prandtl number suppresses the stiffness associated with convection-dominated thermal transport and allows the network to first learn an easier diffusion-dominated temperature field.

A similar continuation is applied to the conductivity ratio appearing in the fluid-solid interface condition:
\begin{equation}
k_{\mathrm{eff}}^*(t)
=
1
+
\left(k_{\mathrm{target}}^* - 1\right)\alpha(t),
\end{equation}
where $k_{\mathrm{target}}^*=250$. Starting from matched conductivities simplifies the conjugate heat-transfer coupling at the interface and gradually restores the target physics as training proceeds.

In addition to parameter continuation, an auxiliary thermal boundary loss is introduced to stabilize the early stage of optimization. \rev{The auxiliary terms at the inlet and top surfaces share a common fading schedule $\beta(t)=\min(t/100{,}000,1)$, which extends beyond the main curriculum.} The inlet loss is assembled as
\begin{equation}
\mathcal{L}_{\mathrm{inlet}}(t)
=
\mathcal{L}_{\mathrm{inlet,scalar}}
+
\revm{\beta(t)}\mathcal{L}_{\mathrm{inlet,flux}}
+
\left[1-\revm{\beta(t)}\right]\mathcal{L}_{\mathrm{inlet,temp}},
\end{equation}
so that training begins with a simpler temperature-value constraint and gradually transitions to the inflow condition. Because the weight of $\mathcal{L}_{\mathrm{inlet,temp}}$ decays to zero, this auxiliary term only guides early optimization and does not alter the final physics.

Similarly, the heated-base loss is defined as
\begin{equation}
\mathcal{L}_{\mathrm{base}}(t)
=
\mathcal{L}_{\mathrm{base},s}
+
\alpha(t)\mathcal{L}_{\mathrm{base},f}.
\end{equation}
This gradual activation improves the thermal coupling between the heated solid base and the fluid region and reduces optimization drift in the coupled thermal problem.

\rev{The top-surface loss is likewise augmented with an auxiliary temperature term during the early stage of training,}
\begin{equation}
\mathcal{L}_{\mathrm{top}}(t)
=
\mathcal{L}_{\mathrm{top}}^{\mathrm{adia}}
+
\left[1-\beta(t)\right]
\mathcal{L}_{\mathrm{top}}^{\mathrm{aux}},
\end{equation}
\rev{which prevents the solid temperature field from collapsing into a trivial flat solution.}

During training, the constant $Pr$ in the fluid-energy residual and the inlet loss is replaced by $Pr_{\mathrm{eff}}(t)$, and the conductivity ratio $k^*$ in the interface loss is replaced by $k_{\mathrm{eff}}^*(t)$.

The total loss is therefore time dependent and can be written as
\begin{equation}
\mathcal{L}_{\mathrm{total}}(t)
=
\sum_{k=1}^{K} w_k^{(t)} \mathcal{L}_k(t),
\end{equation}
where the loss terms include the PDE residuals, the boundary and interface constraints, and the time-dependent inlet, base, and top losses defined above. Because the different loss components can differ substantially in magnitude during training, dynamic gradient normalization is used to balance them \cite{wang_experts_2023}.

Optimization is performed using the SOAP optimizer in its schedule-free variant \cite{defazio_road_2024, vyas_soap_2025, wang_gradient_2025}. SOAP combines matrix preconditioning with Adam-style momentum and bias correction, and has been found to improve the robustness of PINN training for stiff multiphysics problems. In the present implementation, the base learning rate is set to $\eta=10^{-3}$. A linear warm-up is applied over the first 5,000 steps, after which the learning rate decays exponentially by a factor of 0.97 every 2,000 steps during the main stage of training.

The entire framework is implemented in JAX to leverage XLA compilation \cite{bradbury_jax_2018} and GPU execution. A key computational issue in PINNs is the repeated evaluation of first- and second-order spatial derivatives required by the PDE residuals. In the present problem, the network maps a low-dimensional input vector, consisting of spatial coordinates and geometric parameters, to multiple physical fields, which results in a tall Jacobian with respect to the inputs. For this reason, forward-mode automatic differentiation is used for coordinate derivatives, together with reverse-mode backpropagation for parameter updates. This choice is well suited to the present setting and reduces the cost of evaluating the spatial derivatives required by the PDE residuals. The complete training loop, including point sampling, loss evaluation, and optimizer updates, is compiled with just-in-time compiler of JAX to reduce Python overhead. Under the baseline setting, training runs for 200,000 iterations and requires less than four hours on a single NVIDIA RTX 4090D GPU, with a peak memory usage of approximately \qty{20}{\giga\byte}.

\section{Results and discussion}
\label{sec3}

\subsection{Model validation against CFD}
\label{sec3_1}

To assess the predictive accuracy of the proposed PINN model over the entire design space, FEM references are generated using COMSOL Multiphysics 6.3. Because all thermophysical properties were treated as constants, the present flow and heat-transfer problem is effectively one-way coupled: the velocity and pressure fields affect the temperature field, whereas temperature does not feed back into the momentum equation. The references therefore employ a one-way coupled solution strategy, which is fully consistent with the governing physics and avoids unnecessary computational overhead. In this way, the FEM baseline is not artificially disadvantaged in the comparison. Following the grid-independence assessment reported in our previous work on a similar microchannel configuration \cite{wu_jet_2023}, an extra-fine mesh with approximately $2.3\times10^6$ elements is adopted for all FEM cases.

A full-factorial scan is then performed over the parameter space defined in Section~\ref{sec2}. Five values are taken for each of the four geometric parameters, yielding a total of $5^4=625$ FEM simulations. For each parameter combination, 10,000 random evaluation points are sampled in the relevant computational domains, and the reference fields are interpolated onto these locations. In total, this produces $6.25\times10^6$ point-wise comparisons for evaluation of the trained PINN, which is considerably stricter than comparing only global quantities such as pressure drop or peak temperature.

The relative mean squared errors (MSEs) are defined as:
\begin{align}
\text{Error}_u &=
\frac{\frac{1}{N}\sum_{i=1}^{N}\|\mathbf{u}_{\mathrm{pred},i}-\mathbf{u}_{\mathrm{ref},i}\|_2^2}
{\frac{1}{N}\sum_{i=1}^{N}\|\mathbf{u}_{\mathrm{ref},i}\|_2^2+\epsilon}\times100\%, \label{eq:err_u} \\
\text{Error}_T &=
\frac{\frac{1}{N}\sum_{i=1}^{N}(T_{\mathrm{pred},i}-T_{\mathrm{ref},i})^2}
{\frac{1}{N}\sum_{i=1}^{N}T_{\mathrm{ref},i}^2+\epsilon}\times100\%, \label{eq:err_T}
\end{align}
\rev{where \mbox{$N$} is the number of evaluation points for the corresponding field and $\epsilon=10^{-8}$ is introduced to avoid division by zero. Accordingly, the relative \mbox{$L_2$} error is defined as the square root of the relative MSE. The quantitative errors of the full PINN model are summarized in the first row of Table~\mbox{\ref{tab:validation_ablation}}. The trained PINN achieves relative MSEs of 0.66\% and 0.31\% for the velocity and temperature fields, respectively, corresponding to relative \mbox{$L_2$} errors (i.e., the square root of the relative MSEs) of 8.1\% and 5.6\%.}

\begin{figure}
    \centering
    \includegraphics[width=\linewidth]{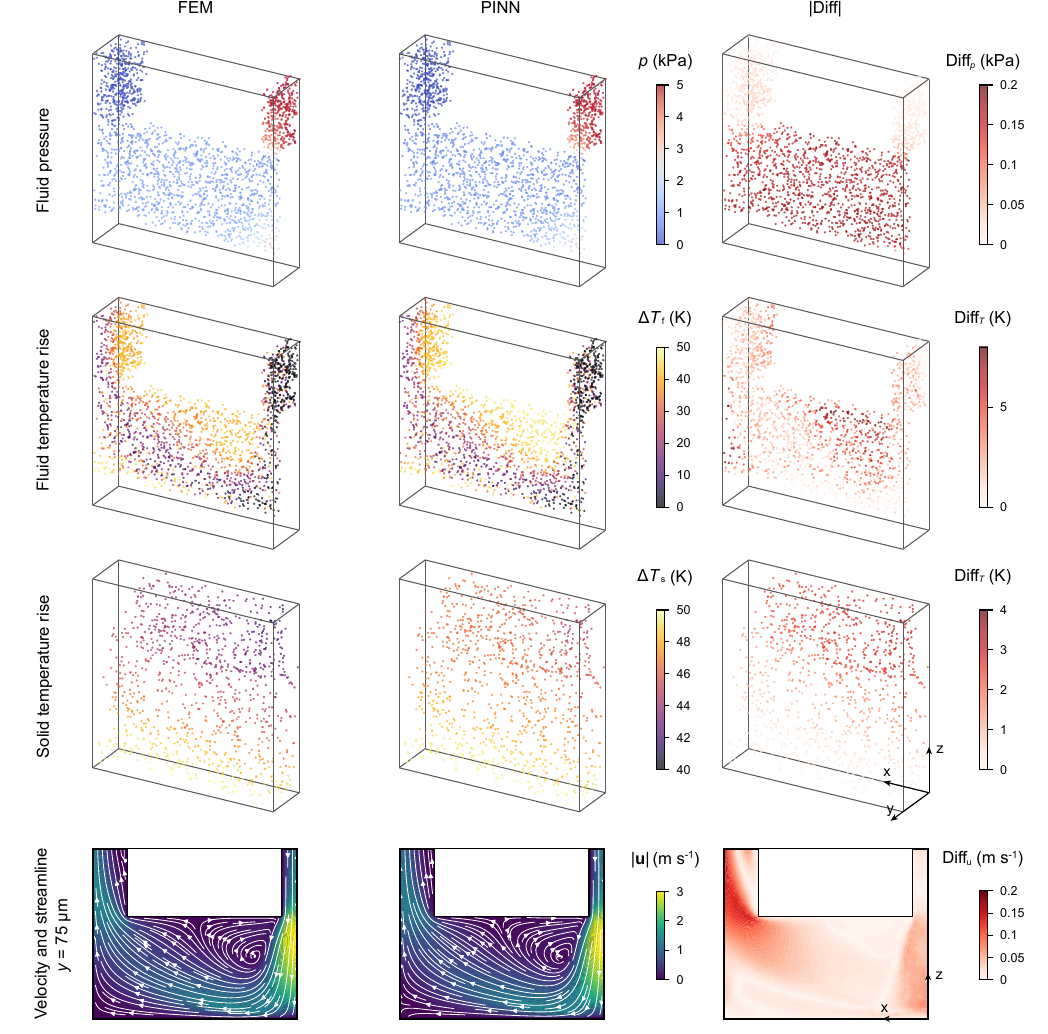}
    \caption{\textbf{Representative three-dimensional field comparison between the FEM reference and the trained parameterized PINN.} This geometry is located at the center of the parameter space, with $W_{\mathrm{in}}=\qty{45}{\micro\meter}$, $W_{\mathrm{out}}=\qty{100}{\micro\meter}$, $W_{\mathrm{mc}}=\qty{50}{\micro\meter}$, and $H_{\mathrm{mc}}=\qty{150}{\micro\meter}$. From left to right, the columns show the FEM solution, the PINN prediction, and their absolute difference. From top to bottom, the rows correspond to fluid pressure $p$, fluid temperature rise $\Delta T_f$, solid temperature rise $\Delta T_s$, fluid velocity magnitude $|\mathbf{u}|$, and the velocity field with streamlines at the $y=\qty{75}{\micro\meter}$ cross-section. The close agreement between the first two columns indicates that the proposed PINN framework faithfully captures the main thermohydraulic features.}
    \label{fig:validation_fields}
\end{figure}

Figure~\ref{fig:validation_fields} further compares the FEM and PINN predictions for a representative geometry with $W_{\mathrm{in}}=\qty{45}{\micro\meter}$, $W_{\mathrm{out}}=\qty{100}{\micro\meter}$, $W_{\mathrm{mc}}=\qty{50}{\micro\meter}$, and $H_{\mathrm{mc}}=\qty{150}{\micro\meter}$. These values lie in the center of the parameter space. Overall, the PINN reproduces the main three-dimensional thermohydraulic features with good point-wise agreement. In the pressure field, it captures both the strong pressure gradient induced by the inlet jet and the local pressure rise in the bottom impingement region where the flow turns into the microchannel section. The predicted fluid-temperature field also matches the FEM reference closely, with the largest discrepancies localized near the outlet. In the solid domain, the model recovers both the dominant through-thickness temperature gradient and the overall heat-spreading pattern in the silicon substrate. Likewise, the velocity-magnitude field is well reproduced, including the localized high-speed regions associated with the inlet jet and the constricted outlet passage. The error maps indicate that the largest pressure discrepancies are mainly concentrated in the microchannel region, whereas the agreement near the inlet and outlet is generally better. Notably, even in the jet inlet region, where the field gradients are strongest, the prediction error remains relatively low. These results confirm that the proposed PINN captures the key local features of the coupled flow and heat transfer fields with good fidelity across regions of both moderate and strong spatial variations.

\begin{figure}
    \centering
    \includegraphics[width=0.5\linewidth]{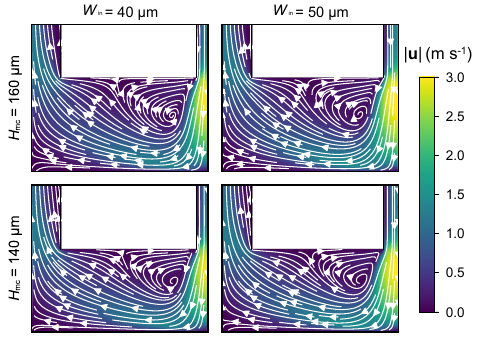}
    \caption{\textbf{Velocity field and streamlines at the $y = \qty{75}{\micro\meter}$ cross-section under different geometric parameter combinations.} The color contour represents the velocity magnitude $|\mathbf{u}|$ in \unit{\meter\per\second}, while the white lines with arrows are the streamlines. The four panels illustrate the flow patterns for different microchannel heights ($H_{\text{mc}} = \qty{160}{\micro\meter}$ and \qty{140}{\micro\meter}, top and bottom rows) and inlet widths ($W_{\text{in}} = \qty{40}{\micro\meter}$ and \qty{50}{\micro\meter}, left and right columns).}
    \label{fig:streamlines}
\end{figure}

Beyond agreement for a single representative case, Fig.~\ref{fig:streamlines} highlights the parametric nature of the learned solution by showing the velocity magnitude and streamlines at the $y=\qty{75}{\micro\meter}$ cross-section for different combinations of $H_{\mathrm{mc}}$ and $W_{\mathrm{in}}$. As the geometry changes, the predicted flow field deforms smoothly with the domain, while remaining fully consistent with the evolving boundaries. In particular, the no-slip walls and channel interfaces are respected across all four configurations, which reflects the effectiveness of the geometric ansatz in embedding geometric constraints into the network input. The influence of the geometric parameters is also physically meaningful. Under the fixed-pressure boundary condition used here, increasing the jet inlet width from $W_{\mathrm{in}}=\qty{40}{\micro\meter}$ to \qty{50}{\micro\meter} reduces the hydraulic resistance and leads to a higher overall flow rate. This trend is clearly captured by the PINN through the expansion of the high-velocity regions, especially the bright yellow acceleration zone near the outlet side. At the same time, the main flow structures are preserved across the parameter variations, including the large recirculating vortex in the microchannel, the low-velocity impingement and turning region near the bottom wall, and the accelerated flow near the jet inlet and outlet. These structures shift and rescale in a physically plausible manner as $H_{\mathrm{mc}}$ and $W_{\mathrm{in}}$ vary, indicating that the model has learned not merely isolated solutions at discrete geometries, but a continuous parameterized mapping between geometry and flow field. It is also worth emphasizing that the trained network accepts the geometric parameters as continuous inputs. Therefore, although validation is carried out on the 625 reference cases, the resulting surrogate is not restricted to these discrete points and can be queried at unseen parameter combinations within the trained design space.

\subsection{Ablation studies}
\label{sec3_2}

The remaining rows of Table~\ref{tab:validation_ablation} compare the full model with several ablated variants. Here, ``Auxiliary loss'' denotes the early-stage thermal boundary continuation introduced in Section~\ref{sec2_4}, where simplified inlet, base, and top-surface temperature constraints are gradually replaced by the full thermal boundary conditions. Overall, removing any one of the proposed components increases the temperature error, indicating that the final performance arises from the combined contributions of geometry-aware embedding, stable optimization, and staged training.

\begin{figure}
    \centering
    \includegraphics[width=0.7\linewidth]{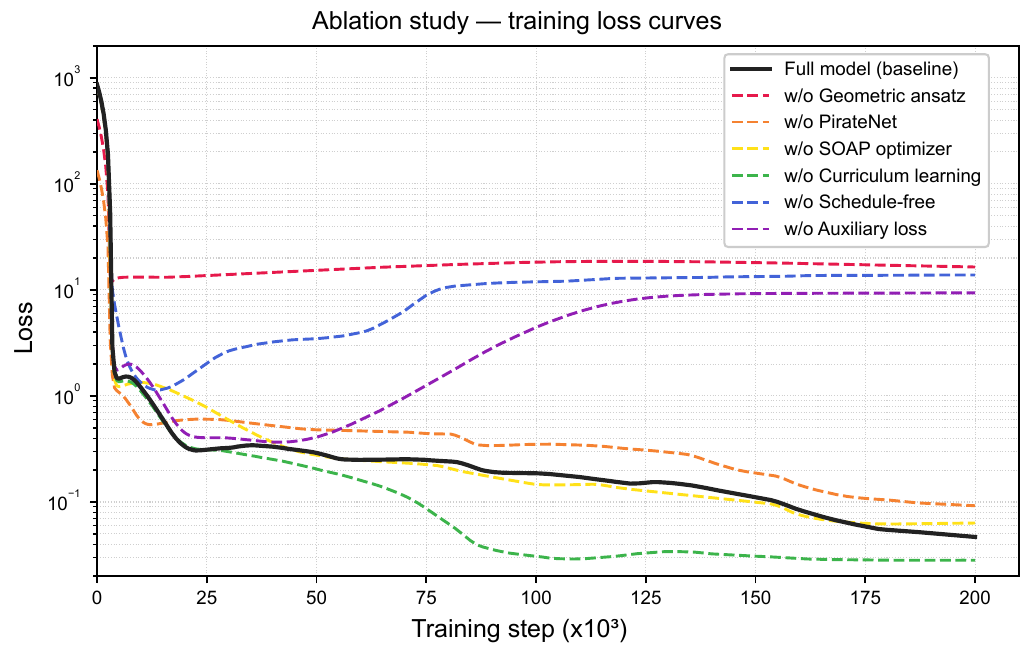}
    \caption{\textbf{Training-loss histories of the full model and the ablation variants.} The full model converges smoothly to a low final loss. The variants without curriculum learning and without SOAP attain training losses comparable to or lower than that of the full model, yet both exhibit higher temperature errors (Table~\ref{tab:validation_ablation}), highlighting that the aggregated training loss alone is insufficient for evaluating the model quality in the present problem.}
    \label{fig:ablation_loss}
\end{figure}

Several important observations can be drawn from Fig.~\ref{fig:ablation_loss}. First, the geometric ansatz is indispensable for parameterized geometries. Once it is removed, the training loss rapidly stagnates at approximately $10^{1}$ and no meaningful convergence is achieved. This behavior is expected: as the geometric parameters vary, the same absolute coordinate may correspond to very different relative positions with respect to the fluid-solid interface and the external boundaries. Without explicit geometric embedding, the network cannot reliably distinguish boundary regions from the interior, and therefore fails to learn the correct solution family.

Second, the schedule-free strategy plays a critical role in optimization stability. Replacing it with a conventional scheduled optimization strategy (denoted as ``w/o Schedule-free'' in the ablation study) results in convergence to a significantly higher loss plateau, preventing the model from reaching the performance of the full configuration. Likewise, removing the auxiliary loss---that is, discarding the transitional terms $\mathcal{L}_{\mathrm{inlet,temp}}$ and $\mathcal{L}_{\mathrm{base},f}$, together with the top-surface temperature term, and imposing the target thermal boundary formulation from the beginning of training---introduces clear late-stage instability. Although the loss initially decreases, it starts to rise again after approximately $4\times10^4$ training steps, suggesting that the auxiliary continuation is important for suppressing optimization drift in this coupled multiphysics problem.

\begin{figure}
    \centering
    \includegraphics[width=\linewidth]{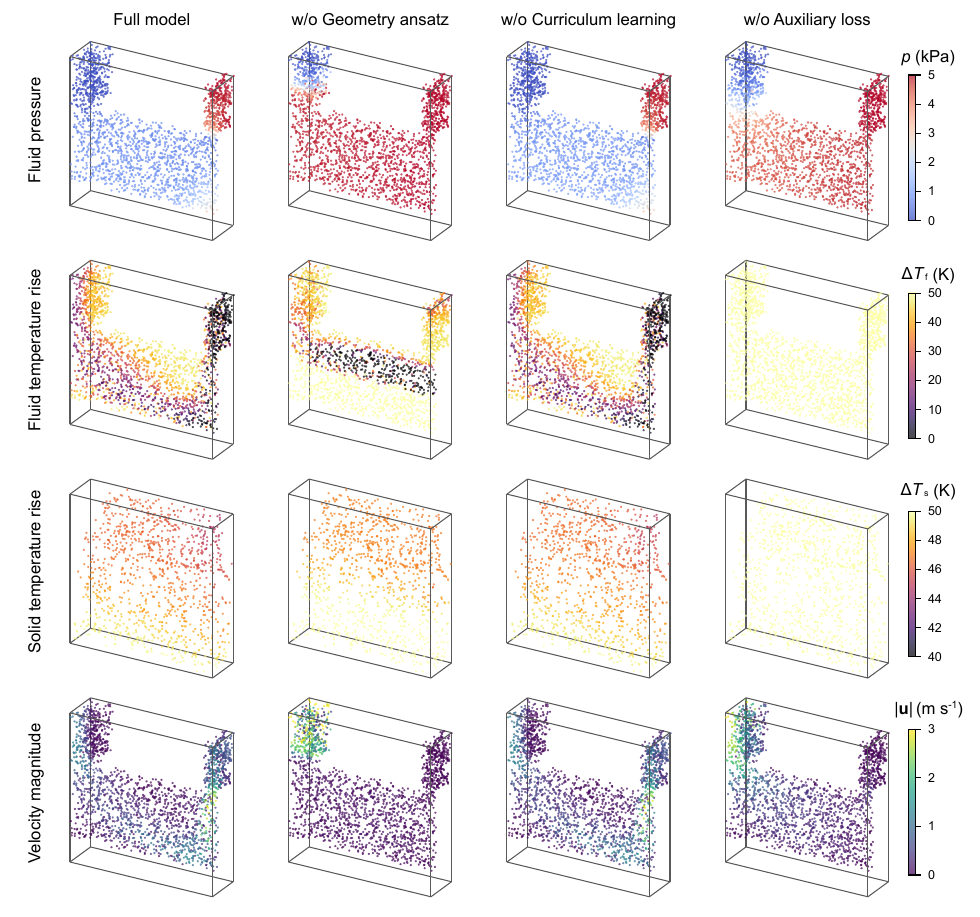}
    \caption{\textbf{Representative field predictions of the baseline model and the ablation variants for a fixed geometry.} From left to right, the columns correspond to the full model, w/o geometric ansatz, w/o curriculum learning, and w/o auxiliary loss. From top to bottom, the rows show the fluid pressure $p$, the fluid temperature rise $\Delta T_\mathrm{f}$, the solid temperature rise $\Delta T_\mathrm{s}$, and the velocity magnitude $|\mathbf{u}|$. }
    \label{fig:ablation_fields}
\end{figure}

The role of curriculum learning is further clarified by the field comparisons in Fig.~\ref{fig:ablation_fields}. As indicated by the green curve in Fig.~\ref{fig:ablation_loss}, the model trained without curriculum learning converges more rapidly and achieves the lowest training loss among all configurations. However, its relative MSE for temperature increases from 0.31\% to 0.39\%. Thus, curriculum learning trades a modest amount of flow-field accuracy for improved temperature fidelity, which is the primary quantity of interest. This result also highlights that, for the present convection-dominated thermohydraulic problem, a lower aggregated PINN loss does not necessarily imply a more accurate physical solution.

The effect of SOAP is also instructive. The run without SOAP reaches a final training loss comparable to that of the full model (Fig.~\ref{fig:ablation_loss}), yet its velocity and temperature errors are markedly larger: removing SOAP increases the relative MSE of velocity from 0.66\% to 1.41\% and the temperature MSE from 0.31\% to 1.49\%, which again underlines that the aggregated training loss alone is not a reliable indicator of field accuracy in the present problem. The pressure field is not emphasized here, since it is directly constrained by the boundary conditions and is therefore less representative of the model's ability to capture the more challenging transport dynamics. By contrast, velocity and especially temperature provide a more meaningful assessment for the present thermohydraulic problem.

More broadly, the ablation study shows that training loss alone is insufficient for evaluating the network quality in this setting. Even when the optimization appears successful in terms of aggregated loss reduction, the predicted physical fields may still be inaccurate, particularly for temperature, which is the stiffest component in the present one-way coupled system. Therefore, field-wise validation against a trusted numerical solver remains essential.

\subsection{Computational efficiency}
\label{sec3_3}

\begin{figure}
    \centering
    \includegraphics[width=0.5\linewidth]{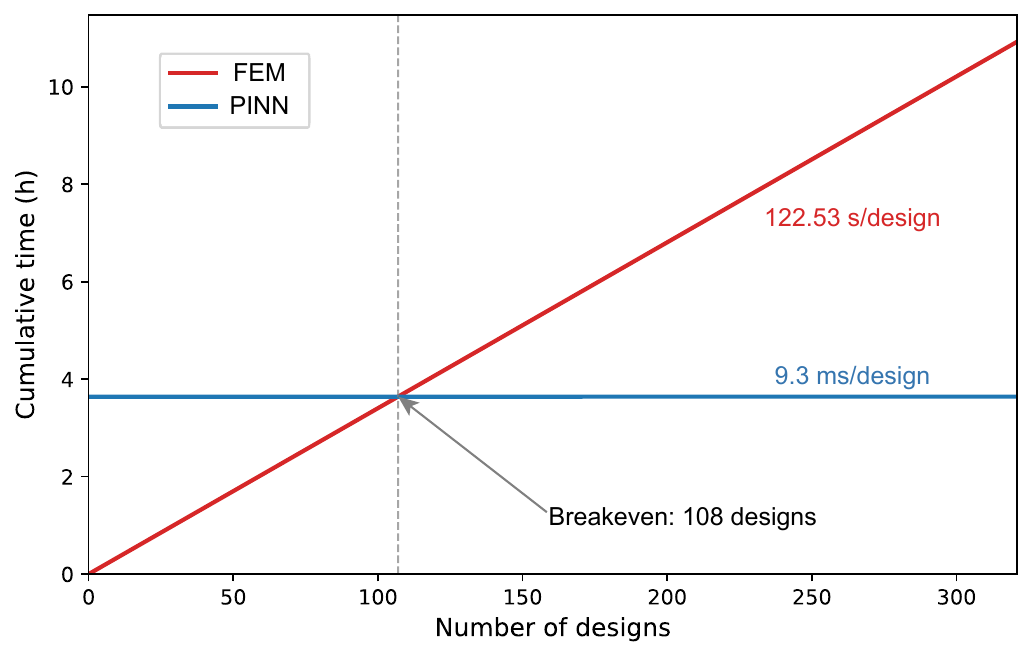}
    \caption{\textbf{Break-even analysis of the cumulative wall-clock time for the FEM and PINN framework.} The red line represents the cumulative cost of the FEM simulations, which increases nearly linearly with the number of evaluated designs. The blue line represents the total cost of the PINN framework, including the one-time training cost and the subsequent inference cost. Because the inference time per design is only \qty{9.3}{\milli\second}, the PINN curve is nearly flat on the scale shown here. The break-even point occurs at 108 design evaluations, after which the PINN-based workflow offers a net wall-clock time saving relative to repeated FEM simulations.}
    \label{fig:breakeven}
\end{figure}

In addition to predictive accuracy, a major motivation for the proposed framework is to reduce the computational cost associated with repeated evaluations over the design space. Table~\ref{tab:computational_cost} summarizes the wall-clock performance and approximate rental-cost estimates of the reference FEM workflow and the proposed PINN surrogate, and Fig.~\ref{fig:breakeven} shows the corresponding cumulative wall-clock break-even behavior.

The reference FEM simulations are carried out on a dual-socket Xeon 8570 node with 112 physical cores in total, with hyper-threading disabled and all 16 memory channels populated. Each simulation uses 56 CPU cores and requires \qty{122.53}{\second} per design point. By contrast, the parameterized PINN is trained on a single NVIDIA RTX 4090D for \qty{3}{\hour} \qty{40}{\minute}, after which the prediction for a new design requires only \qty{9.3}{\milli\second}. This corresponds to a per-design speedup by a factor of approximately $1.32\times10^4$, indicating that the trained network can provide essentially real-time thermohydraulic predictions across the design space.

Because the PINN involves a one-time training cost, its practical benefit is best assessed in terms of the break-even point. Based on wall-clock time, the training cost is amortized after approximately 108 design evaluations. Beyond this point, the cumulative time savings increase nearly linearly with the number of additional queries. This makes the proposed framework particularly attractive for applications that require many repeated evaluations, such as design optimization, sensitivity analysis, and interactive exploration of EMMC geometries.

A similar trend is observed when approximate hardware rental prices are considered. Using representative market rates at the time of submission, about 9 CNY/h for the CPU node and about 2 CNY/h for a single RTX 4090D GPU, the monetary break-even point is reached after only about 24 design evaluations. Nevertheless, wall-clock time is taken as the primary metric in this work, since turnaround time is generally the more relevant constraint in practical engineering design and model development.

Although the FEM and PINN workflows are executed on different hardware platforms, the comparison still reflects their realistic end-to-end computational costs in practical use. In addition, the FEM reference is evaluated in its efficient one-way coupled form, so the reported acceleration is not attributable to an artificially weakened baseline.

\section{Conclusions}
\label{sec4}

In this work, we develop a parameterized physics-informed neural network for the efficient simulation of three-dimensional conjugate heat transfer in embedded manifold microchannels for electronic cooling. The proposed model accepts four geometric parameters together with spatial coordinates as inputs, enabling continuous prediction across a design space of practical interest. Through systematic validation against 625 finite-element simulations, the network achieves relative $L_2$ errors of \rev{8.1\% and 5.6\%} for the velocity and temperature fields, respectively, while reducing the per-design computational time and cost by over four orders of magnitude.

The success of this framework rests on several methodological innovations. A geometric ansatz embedding strategy explicitly encodes the relative position of each point within the parameterized domain, which is essential for training a single network across varying geometries. A curriculum learning scheme that gradually increases the effective Prandtl number and the fluid–solid conductivity ratio stabilizes optimization in the convection-dominated regime. The use of the SOAP optimizer combined with dynamic gradient normalization enables robust convergence of this stiff multiphysics problem. Finally, a JAX-based implementation with forward-mode automatic differentiation and just-in-time compilation reduces the training time to less than four hours on a consumer-grade GPU.

Ablation studies confirm that each of these components contributes significantly to the final accuracy and convergence behavior. Notably, the geometric ansatz is indispensable for parameterized geometries, and curriculum learning helps steer the optimization toward the correct solution branch even when the aggregated loss appears favorable. The trained surrogate model reaches its break-even point after approximately 108 design evaluations in wall-clock time, making it highly attractive for design optimization, sensitivity analysis, and interactive geometry exploration.

Several directions remain for future work. First, extending the parameterized formulation to a wider range of geometric features, including manifold fin shapes and inlet/outlet configurations, would further broaden the practical applicability. Second, incorporating temperature-dependent material properties would address scenarios involving large temperature rises or phase-change cooling. Third, while the current model treats the flow as laminar and steady, extending the framework to turbulent or transient regimes would be valuable for more complex operating conditions. Finally, embedding the trained surrogate within a gradient-based optimization loop could enable rapid, automated design of next-generation embedded microfluidic coolers.

\begin{figure}
    \centering
    \includegraphics[width=\linewidth]{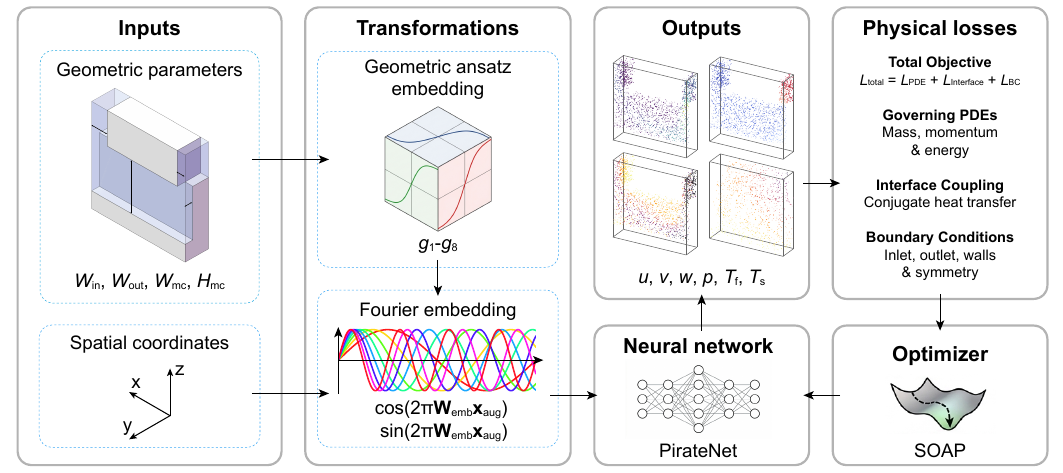}
    \caption{\textbf{Schematic of the proposed parameterized PINN for coupled thermohydraulic prediction in embedded manifold microchannels}. The raw network inputs comprise the spatial coordinates $(x,y,z)$ and four geometric parameters $(W_{\mathrm{in}}, W_{\mathrm{out}}, W_{\mathrm{mc}}, H_{\mathrm{mc}})$. To encode geometric variation and the relative location of each sample point within the parameterized domain, a geometric ansatz embedding generates eight features $(g_1$-$g_8)$, which are concatenated with the original inputs to form a 15-dimensional representation. This representation is further mapped through Fourier feature encoding and passed to the neural network, which predicts the coupled flow and thermal fields $(u,v,w,p,T_{\mathrm{f}},T_{\mathrm{s}})$. The network outputs simultaneously serve as the final solution and as the basis for constructing the physics-informed loss, including PDE residuals, boundary and interface constraints, and auxiliary regularization. The total loss is minimized using the SOAP optimizer.}
    \label{fig:neural_network}
\end{figure}

\begin{table}[p]
\caption{Relative MSEs \rev{and relative $L_2$ errors} with respect to the FEM reference for the full model and for different ablation configurations.}
\label{tab:validation_ablation}
\begin{ruledtabular}
\begin{tabular}{lcccc}
& \multicolumn{2}{c}{Velocity (\%)} & \multicolumn{2}{c}{Temperature (\%)} \\
Configuration & rel. MSE & rel. $L_2$ & rel. MSE & rel. $L_2$ \\
\colrule
Full model (final) & 0.66 & 8.12 & 0.31 & 5.58 \\
w/o Geometric ansatz & 281.33 & 167.73 & 99.91 & 99.95 \\
w/o PirateNet & 3.19 & 17.86 & 0.68 & 8.27 \\
w/o SOAP & 1.41 & 11.86 & 1.49 & 12.19 \\
w/o Curriculum learning & 0.40 & 6.29 & 0.39 & 6.27 \\
w/o Schedule-free & 264.27 & 162.56 & 10.41 & 32.26 \\
w/o Auxiliary loss & 128.15 & 113.20 & 10.41 & 32.27 \\
\end{tabular}
\end{ruledtabular}
\end{table}

\begin{table}[p]
\caption{Wall-clock performance and approximate rental-cost comparison between the reference FEM workflow and the proposed PINN surrogate model.}
\label{tab:computational_cost}
\begin{ruledtabular}
\begin{tabular}{lcc}
Metric & FEM & PINN \\
\colrule
\multicolumn{3}{l}{\textit{Cost basis}} \\
Approx. rental price per hour & 9 CNY/h & 2 CNY/h \\[2pt]
\multicolumn{3}{l}{\textit{Wall-clock performance}} \\
One-time setup / training time & -- & \qty{3.67}{\hour} \\
Runtime per design & \qty{122.53}{\second} & \qty{9.3}{\milli\second} \\
Per-design speedup & -- & $\sim 1.32\times10^4$ \\
Time break-even No. of designs & -- & 108 \\
Price break-even No. of designs & -- & 24 \\
\end{tabular}
\end{ruledtabular}
\end{table}

\clearpage
\section*{Conflict of interest}
The authors declare that they have no known competing financial interests or personal relationships that could have appeared to influence the work reported in this paper.

\begin{acknowledgments}
This work was supported by the Science Fund for Creative Research Groups from the National Natural Science Foundation of China (No. 52521007), the Scientific Research Innovation Capability Support Project for Young Faculty (ZYGXQNJSKYCXNLZCXM-E1) from the Ministry of Education of China, the Beijing Natural Science Foundation (QY23035), the National Key R \& D Project from the Ministry of Science and Technology of China (No. 2022YFA1203100 and No. 2024YFA1207900), and the High-performance Computing Platform of Peking University. We thank Haohao Gu and Sensen He from the Baidu Frontier Research Department for their valuable discussions. B.S. acknowledges support from the New Cornerstone Science Foundation through the XPLORER PRIZE.
\end{acknowledgments}

\bibliography{references}

@inproceedings{hennigh_nvidia_2021,
	address = {Cham},
	series = {Lecture {Notes} in {Computer} {Science}},
	title = {{NVIDIA} {SimNet}™: {An} {AI}-{Accelerated} {Multi}-{Physics} {Simulation} {Framework}},
	isbn = {978-3-030-77977-1},
	shorttitle = {{NVIDIA} {SimNet}™},
	doi = {10.1007/978-3-030-77977-1\_36},
	booktitle = {Computational {Science} – {ICCS} 2021},
	publisher = {Springer International Publishing},
	author = {Hennigh, Oliver and Narasimhan, Susheela and Nabian, Mohammad Amin and Subramaniam, Akshay and Tangsali, Kaustubh and Fang, Zhiwei and Rietmann, Max and Byeon, Wonmin and Choudhry, Sanjay},
	editor = {Paszynski, Maciej and Kranzlmüller, Dieter and Krzhizhanovskaya, Valeria V. and Dongarra, Jack J. and Sloot, Peter M.A.},
	year = {2021},
	pages = {447--461},
}

@article{raissi_physics-informed_2019,
	title = {Physics-informed neural networks: {A} deep learning framework for solving forward and inverse problems involving nonlinear partial differential equations},
	volume = {378},
	issn = {0021-9991},
	shorttitle = {Physics-informed neural networks},
	url = {https://www.sciencedirect.com/science/article/pii/S0021999118307125},
	doi = {10.1016/j.jcp.2018.10.045},
	urldate = {2023-07-18},
	journal = {Journal of Computational Physics},
	author = {Raissi, M. and Perdikaris, P. and Karniadakis, G. E.},
	month = feb,
	year = {2019},
	pages = {686--707},
}

@article{van_erp_co-designing_2020,
	title = {Co-designing electronics with microfluidics for more sustainable cooling},
	volume = {585},
	issn = {0028-0836, 1476-4687},
	url = {https://www.nature.com/articles/s41586-020-2666-1},
	doi = {10.1038/s41586-020-2666-1},
	number = {7824},
	urldate = {2023-07-12},
	journal = {Nature},
	author = {Van Erp, Remco and Soleimanzadeh, Reza and Nela, Luca and Kampitsis, Georgios and Matioli, Elison},
	month = sep,
	year = {2020},
	pages = {211--216},
}

@article{bar-cohen_icecool_2021,
	title = {The {ICECool} {Fundamentals} {Effort} on {Evaporative} {Cooling} of {Microelectronics}},
	volume = {11},
	issn = {2156-3950, 2156-3985},
	url = {https://ieeexplore.ieee.org/document/9536673/},
	doi = {10.1109/TCPMT.2021.3111114},
	number = {10},
	urldate = {2023-07-12},
	journal = {IEEE Transactions on Components, Packaging and Manufacturing Technology},
	author = {Bar-Cohen, Avram and Asheghi, Mehdi and Chainer, Timothy J. and Garimella, Suresh V. and Goodson, Kenneth and Gorle, Catherine and Mandel, Raphael and Maurer, Joseph J. and Ohadi, Michael and Palko, James W. and Parida, Pritish R. and Peles, Yoav and Plawsky, Joel L. and Schultz, Mark D. and Weibel, Justin A. and Joshi, Yogendra},
	month = oct,
	year = {2021},
	pages = {1546--1564},
}

@article{bar-cohen_embedded_2019,
	title = {Embedded {Cooling} for {Wide} {Bandgap} {Power} {Amplifiers}: {A} {Review}},
	volume = {141},
	issn = {1043-7398, 1528-9044},
	shorttitle = {Embedded {Cooling} for {Wide} {Bandgap} {Power} {Amplifiers}},
	url = {https://asmedigitalcollection.asme.org/electronicpackaging/article/doi/10.1115/1.4043404/726543/Embedded-Cooling-for-Wide-Bandgap-Power-Amplifiers},
	doi = {10.1115/1.4043404},
	number = {4},
	urldate = {2023-07-12},
	journal = {Journal of Electronic Packaging},
	author = {Bar-Cohen, A. and Maurer, J. J. and Altman, D. H.},
	month = dec,
	year = {2019},
	pages = {040803},
}

@misc{cuomo_scientific_2022,
	title = {Scientific {Machine} {Learning} through {Physics}-{Informed} {Neural} {Networks}: {Where} we are and {What}'s next},
	shorttitle = {Scientific {Machine} {Learning} through {Physics}-{Informed} {Neural} {Networks}},
	url = {http://arxiv.org/abs/2201.05624},
	urldate = {2023-07-17},
	publisher = {arXiv},
	author = {Cuomo, Salvatore and di Cola, Vincenzo Schiano and Giampaolo, Fabio and Rozza, Gianluigi and Raissi, Maziar and Piccialli, Francesco},
	month = jun,
	year = {2022},
	note = {arXiv:2201.05624 [physics]},
}

@article{raissi_hidden_2020,
	title = {Hidden fluid mechanics: {Learning} velocity and pressure fields from flow visualizations},
	volume = {367},
	shorttitle = {Hidden fluid mechanics},
	url = {https://www.science.org/doi/10.1126/science.aaw4741},
	doi = {10.1126/science.aaw4741},
	number = {6481},
	urldate = {2024-04-22},
	journal = {Science},
	publisher = {American Association for the Advancement of Science},
	author = {Raissi, Maziar and Yazdani, Alireza and Karniadakis, George Em},
	month = feb,
	year = {2020},
	pages = {1026--1030},
}

@article{wu_jet_2023,
	title = {Jet microchannel with sawtooth wall for efficient cooling of high-power electronics},
	volume = {206},
	issn = {0017-9310},
	url = {https://www.sciencedirect.com/science/article/pii/S0017931023001102},
	doi = {10.1016/j.ijheatmasstransfer.2023.123955},
	urldate = {2023-11-05},
	journal = {International Journal of Heat and Mass Transfer},
	author = {Wu, Zhihu and Jiang, Zhiyao and Yan, Wujuan and Yang, Yuchi and Kang, Jin and Zheng, Kai and Bu, Weihai and Wang, Wei and Song, Bai},
	month = jun,
	year = {2023},
	pages = {123955},
}

@article{baydin_automatic_2018,
	title = {Automatic {Differentiation} in {Machine} {Learning}: a {Survey}},
	volume = {18},
	issn = {1533-7928},
	shorttitle = {Automatic {Differentiation} in {Machine} {Learning}},
	url = {http://jmlr.org/papers/v18/17-468.html},
	number = {153},
	urldate = {2024-04-26},
	journal = {Journal of Machine Learning Research},
	author = {Baydin, Atilim Gunes and Pearlmutter, Barak A. and Radul, Alexey Andreyevich and Siskind, Jeffrey Mark},
	year = {2018},
	pages = {1--43},
}

@article{wu_efficient_2024,
	title = {Efficient thermal management of high-power electronics via jet-enhanced {HU}-type manifold microchannel},
	volume = {221},
	issn = {0017-9310},
	url = {https://www.sciencedirect.com/science/article/pii/S0017931023012589},
	doi = {10.1016/j.ijheatmasstransfer.2023.125113},
	urldate = {2024-02-26},
	journal = {International Journal of Heat and Mass Transfer},
	author = {Wu, Zhihu and Xiao, Wei and Song, Bai},
	month = apr,
	year = {2024},
	pages = {125113},
}

@inproceedings{rahaman_spectral_2019,
	title = {On the {Spectral} {Bias} of {Neural} {Networks}},
	issn = {2640-3498},
	url = {https://proceedings.mlr.press/v97/rahaman19a.html},
	urldate = {2024-04-26},
	booktitle = {Proceedings of the 36th {International} {Conference} on {Machine} {Learning}},
	publisher = {PMLR},
	author = {Rahaman, Nasim and Baratin, Aristide and Arpit, Devansh and Draxler, Felix and Lin, Min and Hamprecht, Fred and Bengio, Yoshua and Courville, Aaron},
	month = may,
	year = {2019},
	pages = {5301--5310},
}

@article{li_computational_2024,
	title = {A computational framework for neural network-based variational {Monte} {Carlo} with {Forward} {Laplacian}},
	volume = {6},
	copyright = {2024 The Author(s), under exclusive licence to Springer Nature Limited},
	issn = {2522-5839},
	url = {https://www.nature.com/articles/s42256-024-00794-x},
	doi = {10.1038/s42256-024-00794-x},
	number = {2},
	urldate = {2024-03-12},
	journal = {Nature Machine Intelligence},
	publisher = {Nature Publishing Group},
	author = {Li, Ruichen and Ye, Haotian and Jiang, Du and Wen, Xuelan and Wang, Chuwei and Li, Zhe and Li, Xiang and He, Di and Chen, Ji and Ren, Weiluo and Wang, Liwei},
	month = feb,
	year = {2024},
	pages = {209--219},
}

@misc{krishnapriyan_characterizing_2021,
	title = {Characterizing possible failure modes in physics-informed neural networks},
	url = {http://arxiv.org/abs/2109.01050},
	doi = {10.48550/arXiv.2109.01050},
	urldate = {2025-08-26},
	publisher = {arXiv},
	author = {Krishnapriyan, Aditi S. and Gholami, Amir and Zhe, Shandian and Kirby, Robert M. and Mahoney, Michael W.},
	month = nov,
	year = {2021},
	note = {arXiv:2109.01050 [cs]},
}

@article{yang_parameterized_2025,
	title = {A {Parameterized} {Thermal} {Simulation} {Method} {Based} on {Physics}-{Informed} {Neural} {Networks} for {Fast} {Power} {Module} {Thermal} {Design}},
	volume = {40},
	issn = {1941-0107},
	url = {https://ieeexplore.ieee.org/document/10909630},
	doi = {10.1109/TPEL.2025.3547390},
	number = {7},
	urldate = {2025-07-29},
	journal = {IEEE Transactions on Power Electronics},
	author = {Yang, Yayong and Wang, Zhiqiang and Liao, Yu and Kong, Wubin and Shi, Xiaojie and Hu, Run and Yao, Yonggang},
	month = jul,
	year = {2025},
	pages = {9200--9210},
}

@misc{wang_simulating_2025,
	title = {Simulating {Three}-dimensional {Turbulence} with {Physics}-informed {Neural} {Networks}},
	url = {http://arxiv.org/abs/2507.08972},
	doi = {10.48550/arXiv.2507.08972},
	urldate = {2025-07-27},
	publisher = {arXiv},
	author = {Wang, Sifan and Sankaran, Shyam and Stinis, Panos and Perdikaris, Paris},
	month = jul,
	year = {2025},
	note = {arXiv:2507.08972 [cs]},
}

@misc{vyas_soap_2025,
	title = {{SOAP}: {Improving} and {Stabilizing} {Shampoo} using {Adam}},
	shorttitle = {{SOAP}},
	url = {http://arxiv.org/abs/2409.11321},
	doi = {10.48550/arXiv.2409.11321},
	urldate = {2025-07-27},
	publisher = {arXiv},
	author = {Vyas, Nikhil and Morwani, Depen and Zhao, Rosie and Kwun, Mujin and Shapira, Itai and Brandfonbrener, David and Janson, Lucas and Kakade, Sham},
	month = jan,
	year = {2025},
	note = {arXiv:2409.11321 [cs]},
}

@misc{wang_random_2022,
	title = {Random {Weight} {Factorization} {Improves} the {Training} of {Continuous} {Neural} {Representations}},
	url = {http://arxiv.org/abs/2210.01274},
	doi = {10.48550/arXiv.2210.01274},
	urldate = {2025-07-07},
	publisher = {arXiv},
	author = {Wang, Sifan and Wang, Hanwen and Seidman, Jacob H. and Perdikaris, Paris},
	month = oct,
	year = {2022},
	note = {arXiv:2210.01274 [cs]},
}

@misc{wang_experts_2023,
	title = {An {Expert}'s {Guide} to {Training} {Physics}-informed {Neural} {Networks}},
	url = {http://arxiv.org/abs/2308.08468},
	doi = {10.48550/arXiv.2308.08468},
	urldate = {2025-06-19},
	publisher = {arXiv},
	author = {Wang, Sifan and Sankaran, Shyam and Wang, Hanwen and Perdikaris, Paris},
	month = aug,
	year = {2023},
	note = {arXiv:2308.08468 [cs]},
}

@misc{wang_piratenets_2024,
	title = {{PIRATENETS}: {PHYSICS}-{INFORMED} {DEEP} {LEARNING} {WITH}  {RESIDUAL} {ADAPTIVE} {NETWORKS}},
	shorttitle = {{PirateNets}},
	url = {http://arxiv.org/abs/2402.00326},
	doi = {10.48550/arXiv.2402.00326},
	urldate = {2025-04-26},
	publisher = {arXiv},
	author = {Wang, Sifan and Li, Bowen and Chen, Yuhan and Perdikaris, Paris},
	month = feb,
	year = {2024},
	note = {arXiv:2402.00326 [cs]},
}

@article{zhang_physics-informed_2024,
	title = {Physics-informed neural network simulation of conjugate heat transfer in manifold microchannel heat sinks for high-power {IGBT} cooling},
	volume = {159},
	issn = {0735-1933},
	url = {https://www.sciencedirect.com/science/article/pii/S073519332400798X},
	doi = {10.1016/j.icheatmasstransfer.2024.108036},
	urldate = {2024-09-10},
	journal = {International Communications in Heat and Mass Transfer},
	author = {Zhang, Xiangzhi and Tu, Chaofan and Yan, Yuying},
	month = dec,
	year = {2024},
	pages = {108036},
}

@article{wu_jet-enhanced_2025,
	title = {Jet-enhanced manifold microchannels for cooling electronics up to a heat flux of 3,000 {W} cm-2},
	volume = {8},
	copyright = {2025 The Author(s), under exclusive licence to Springer Nature Limited},
	issn = {2520-1131},
	url = {https://www.nature.com/articles/s41928-025-01449-4},
	doi = {10.1038/s41928-025-01449-4},
	number = {9},
	urldate = {2026-01-08},
	journal = {Nature Electronics},
	publisher = {Nature Publishing Group},
	author = {Wu, Zhihu and Xiao, Wei and He, Haiyu and Wang, Wei and Song, Bai},
	month = sep,
	year = {2025},
	pages = {810--817},
}

@misc{bradbury_jax_2018,
	title = {{JAX}: composable transformations of {Python}+{NumPy} programs},
	url = {http://github.com/jax-ml/jax},
	author = {Bradbury, James and Frostig, Roy and Hawkins, Peter and Johnson, Matthew James and Leary, Chris and Maclaurin, Dougal and Necula, George and Paszke, Adam and VanderPlas, Jake and Wanderman-Milne, Skye and Zhang, Qiao},
	year = {2018},
}

@misc{deepmind_deepmind_2020,
	title = {The {DeepMind} {JAX} {Ecosystem}},
	url = {http://github.com/google-deepmind},
	author = {{DeepMind} and Babuschkin, Igor and Baumli, Kate and Bell, Alison and Bhupatiraju, Surya and Bruce, Jake and Buchlovsky, Peter and Budden, David and Cai, Trevor and Clark, Aidan and Danihelka, Ivo and Dedieu, Antoine and Fantacci, Claudio and Godwin, Jonathan and Jones, Chris and Hemsley, Ross and Hennigan, Tom and Hessel, Matteo and Hou, Shaobo and Kapturowski, Steven and Keck, Thomas and Kemaev, Iurii and King, Michael and Kunesch, Markus and Martens, Lena and Merzic, Hamza and Mikulik, Vladimir and Norman, Tamara and Papamakarios, George and Quan, John and Ring, Roman and Ruiz, Francisco and Sanchez, Alvaro and Sartran, Laurent and Schneider, Rosalia and Sezener, Eren and Spencer, Stephen and Srinivasan, Srivatsan and Stanojević, Miloš and Stokowiec, Wojciech and Wang, Luyu and Zhou, Guangyao and Viola, Fabio},
	year = {2020},
}

@misc{defazio_road_2024,
	title = {The {Road} {Less} {Scheduled}},
	author = {Defazio, Aaron and Yang, Xingyu and Mehta, Harsh and Mishchenko, Konstantin and Khaled, Ahmed and Cutkosky, Ashok},
	year = {2024},
	note = {\_eprint: 2405.15682},
}

@inproceedings{bengio_curriculum_2009,
	address = {New York, NY, USA},
	series = {{ICML} '09},
	title = {Curriculum learning},
	isbn = {978-1-60558-516-1},
	url = {https://doi.org/10.1145/1553374.1553380},
	doi = {10.1145/1553374.1553380},
	urldate = {2026-01-08},
	booktitle = {Proceedings of the 26th {Annual} {International} {Conference} on {Machine} {Learning}},
	publisher = {Association for Computing Machinery},
	author = {Bengio, Yoshua and Louradour, Jérôme and Collobert, Ronan and Weston, Jason},
	month = jun,
	year = {2009},
	pages = {41--48},
}

@article{tuckerman_high-performance_1981,
	title = {High-performance heat sinking for {VLSI}},
	volume = {2},
	issn = {1558-0563},
	url = {https://ieeexplore.ieee.org/document/1481851},
	doi = {10.1109/EDL.1981.25367},
	number = {5},
	urldate = {2026-01-08},
	journal = {IEEE Electron Device Letters},
	author = {Tuckerman, D.B. and Pease, R.F.W.},
	month = may,
	year = {1981},
	pages = {126--129},
}

@article{escher_experimental_2010,
	title = {Experimental {Investigation} of an {Ultrathin} {Manifold} {Microchannel} {Heat} {Sink} for {Liquid}-{Cooled} {Chips}},
	volume = {132},
	issn = {0022-1481},
	url = {https://doi.org/10.1115/1.4001306},
	doi = {10.1115/1.4001306},
	number = {081402},
	urldate = {2026-01-08},
	journal = {Journal of Heat Transfer},
	author = {Escher, W. and Brunschwiler, T. and Michel, B. and Poulikakos, D.},
	month = jun,
	year = {2010},
}

@article{lin_single-phase_2021,
	title = {Single-phase and {Two}-phase {Flow} and {Heat} {Transfer} in {Microchannel} {Heat} {Sink} with {Various} {Manifold} {Arrangements}},
	volume = {171},
	issn = {0017-9310},
	url = {https://www.sciencedirect.com/science/article/pii/S0017931021002210},
	doi = {10.1016/j.ijheatmasstransfer.2021.121118},
	urldate = {2026-01-08},
	journal = {International Journal of Heat and Mass Transfer},
	author = {Lin, Yuhao and Luo, Yang and Li, Wei and Cao, Yanlong and Tao, Zhi and Shih, Tom I-P.},
	month = jun,
	year = {2021},
	pages = {121118},
}

@article{jiang_synergistic_2026,
	title = {Synergistic thermohydraulic design and performance limit of embedded microfluidic chip cooling},
	volume = {283},
	issn = {1359-4311},
	url = {https://www.sciencedirect.com/science/article/pii/S1359431125035082},
	doi = {10.1016/j.applthermaleng.2025.128916},
	urldate = {2026-01-08},
	journal = {Applied Thermal Engineering},
	author = {Jiang, Zhiyao and Xiao, Wei and Li, Weiheng and Wu, Zhihu and Wang, Wei and Song, Bai},
	month = jan,
	year = {2026},
	pages = {128916},
}

@article{zhang_survey_2021,
	title = {A survey on data center cooling systems: {Technology}, power consumption modeling and control strategy optimization},
	volume = {119},
	issn = {1383-7621},
	shorttitle = {A survey on data center cooling systems},
	url = {https://www.sciencedirect.com/science/article/pii/S1383762121001739},
	doi = {10.1016/j.sysarc.2021.102253},
	urldate = {2026-01-09},
	journal = {Journal of Systems Architecture},
	author = {Zhang, Qingxia and Meng, Zihao and Hong, Xianwen and Zhan, Yuhao and Liu, Jia and Dong, Jiabao and Bai, Tian and Niu, Junyu and Deen, M. Jamal},
	month = oct,
	year = {2021},
	pages = {102253},
}

@article{fabbri_state_2007,
	title = {State of the {Art} of {High} {Heat} {Flux} {Cooling} {Technologies}},
	volume = {28},
	doi = {10.1080/01457630601117799},
	journal = {Heat Transfer Engineering},
	author = {Fabbri, Matteo and Park, Jung and Wojtan, Leszek and Thome, John and Michel, Bruno},
	month = apr,
	year = {2007},
}

@misc{noauthor_physicsnemo_nodate,
	title = {{PhysicsNeMo} {Sym} — {NVIDIA} {PhysicsNeMo} {Framework}},
	url = {https://docs.nvidia.com/physicsnemo/latest/physicsnemo-sym/index.html},
	urldate = {2026-01-12},
}

@article{wang_gradient_2025,
  title={Gradient Alignment in Physics-informed Neural Networks: A Second-Order Optimization Perspective},
  author={Wang, Sifan and Bhartari, Ananyae Kumar and Li, Bowen and Perdikaris, Paris},
  journal={arXiv preprint arXiv:2502.00604},
  year={2025}
}

@article{yu_general_2023,
	title = {General deep learning framework for emissivity engineering},
	journal = {Light: Science \& Applications},
	author = {Yu, S. and Zhou, P. and Xi, W. and others},
	year = {2023},
	volume = {12},
	pages = {291},
}

@article{hu_machine-learning-optimized_2020,
	title = {Machine-learning-optimized aperiodic superlattice minimizes coherent phonon heat conduction},
	journal = {Physical Review X},
	author = {Hu, R. and Iwamoto, S. and Feng, L. and Ju, S. and Hu, S. and Ohnishi, M. and Nagai, N. and Hirakawa, K. and Shiomi, J.},
	year = {2020},
	volume = {10},
	pages = {021050},
}

@article{hu_machine_2020,
	title = {Machine learning-optimized {Tamm} emitter for high-performance thermophotovoltaic system with detailed balance analysis},
	journal = {Nano Energy},
	author = {Hu, R. and Song, J. and Liu, Y. and Xi, W. and Zhao, Y. and Yu, X. and Cheng, Q. and Tao, G. and Luo, X.},
	year = {2020},
	volume = {72},
	pages = {104687},
}

@article{yu_ultrahigh_2024,
	title = {Ultrahigh visible-transparency, submicrometer, and polymer-free radiative cooling meta-glass coating for building energy saving},
	journal = {ACS Photonics},
	author = {Yu, S. and Yu, J.-S. and Chen, Z. and Li, Q. and Wang, Z. and Luo, X. and Kim, S.-K. and Hu, R.},
	year = {2024},
	volume = {11},
	number = {8},
	pages = {3412--3423},
}

@article{chen_strong_2025,
	title = {Strong nonreciprocal broadband thermal radiation via materials informatics inverse design},
	journal = {Advanced Optical Materials},
	author = {Chen, Z. and Hu, R.},
	year = {2025},
	volume = {13},
	pages = {2401219},
}

@article{li_biomimetic_2026,
	title = {Biomimetic fish-fin microchannels for energy-efficient manifold liquid cooling},
	journal = {Applied Thermal Engineering},
	author = {Li, H. and Wang, Z. and Choi, W. and Hu, R.},
	year = {2026},
	volume = {304},
	pages = {132654},
}

@article{jiang_embedded_2026,
	title = {Embedded micro-pin-fin heat sink with {3D} manifold for cooling high-power electronics},
	journal = {International Journal of Heat and Mass Transfer},
	author = {Jiang, Z. and Xiao, W. and Li, W. and Song, B.},
	year = {2026},
	volume = {273},
	pages = {129613},
}

@article{yang_vibe-fdtr_2026,
	title = {Vibe-{FDTR}: {An} agent-oriented framework for reproducible frequency-domain thermoreflectance data analysis},
	journal = {Advanced Scientific Instruments},
	author = {Yang, F. and Li, W. and Song, B.},
	year = {2026},
	pages = {100049},
}

@article{xiao_horizontal-up_2026,
	title = {Horizontal-up microfluidics for embedded chip cooling with a dielectric fluid},
	journal = {Nature Electronics},
	author = {Xiao, W. and Wu, Z. and Jiang, Z. and Wang, Y. and Wang, W. and Song, B.},
	year = {2026},
	pages = {1--8},
}

@article{wu_correcting_2024,
	title = {Correcting force error-induced underestimation of lattice thermal conductivity in machine learning molecular dynamics},
	volume = {161},
	issn = {0021-9606},
	url = {https://doi.org/10.1063/5.0213811},
	doi = {10.1063/5.0213811},
	number = {1},
	urldate = {2026-10-29},
	journal = {The Journal of Chemical Physics},
	author = {Wu, Xiguang and Zhou, Wenjiang and Dong, Haikuan and Ying, Penghua and Wang, Yanzhou and Song, Bai and Fan, Zheyong and Xiong, Shiyun},
	month = jul,
	year = {2024},
	pages = {014103},
}

@article{zhou_insight_2025,
	title = {Insight into the effect of force error on the thermal conductivity from machine-learned potentials},
	volume = {50},
	issn = {2542-5293},
	url = {https://www.sciencedirect.com/science/article/pii/S2542529324003146},
	doi = {10.1016/j.mtphys.2024.101638},
	urldate = {2026-10-29},
	journal = {Materials Today Physics},
	author = {Zhou, Wenjiang and Liang, Nianjie and Wu, Xiguang and Xiong, Shiyun and Fan, Zheyong and Song, Bai},
	month = jan,
	year = {2025},
	pages = {101638},
}

\end{document}